\documentclass[pre,amsmath]{revtex4}
\usepackage{graphicx,url,hyperref} 
\usepackage{color}
\begin{document}

\title{Equilibration in the Nos\'{e}-Hoover isokinetic ensemble: Effect of inter-particle interactions}
\author{Shamik Gupta$^{1}$ and Stefano Ruffo$^{2}$}
\affiliation{ $^{1}$Department of Physics, Ramakrishna Mission
Vivekananda University, Belur Math, Howrah 711202, India;
shamik.gupta@rkmvu.ac.in\\ $^{2}$SISSA, INFN and ISC-CNR - Via Bonomea
265, I-34136 Trieste, Italy; ruffo@sissa.it
}
\begin{abstract}
We investigate the stationary and dynamic properties of the celebrated Nos\'{e}-Hoover
dynamics of many-body interacting Hamiltonian systems, with an emphasis on the effect of inter-particle
interactions. To this end, we consider a model system with both short-
and long-range interactions. The Nos\'{e}-Hoover dynamics aims to generate the canonical equilibrium
distribution of a system at a desired temperature by employing a set of 
time-reversible, deterministic equations of motion. A signature
of canonical equilibrium is a single-particle momentum distribution
that is Gaussian. We find that the equilibrium properties of the
system within the Nos\'{e}-Hoover dynamics coincides with that within
the canonical ensemble. Moreover, starting from out-of-equilibrium initial
conditions, the average kinetic energy of the system
relaxes to its target
value over a {\it size-independent} timescale. However, quite
surprisingly, our results indicate that under the same conditions and
with only long-range interactions present in the system, the momentum distribution relaxes to its Gaussian form in equilibrium over a scale that {\it
diverges with the system size}. On adding short-range interactions, the
relaxation is found to occur over a timescale that has a much weaker dependence on
system size. This system-size dependence of the timescale vanishes when only short-range
interactions are present in the system. An implication of such an ultra-slow relaxation when
only long-range interactions are present in the system is that
macroscopic observables other than the average kinetic energy when
estimated in the Nos\'{e}-Hoover dynamics may take an unusually
long time to relax to its canonical equilibrium value. Our work underlines the
crucial role that interactions play in deciding the equivalence between
Nos\'{e}-Hoover and canonical equilibrium. \\
{\bf Invited contribution to Entropy {\em Special Issue}
entitled \href{http://www.mdpi.com/journal/entropy/special_issues/small_systems}{Thermodynamics and Statistical Mechanics of Small Systems}: Academic Editor: {\bf Andrea Puglisi, Alessandro Sarracino,
and Angelo Vulpiani}}
\end{abstract}
\maketitle

\section{Introduction}
\label{sec:introduction}

Often, one needs in studies in nonlinear dynamics and statistical physics to investigate the
dynamical properties of a many-body interacting Hamiltonian system evolving
under the condition of a constant temperature. For example,
one might be interested in studying the dynamical properties of the system
in canonical equilibrium at a certain temperature $T$, with the
temperature being proportional to the average kinetic energy of the system by virtue of
the Theorem of Equipartition (In this
work, we measure temperatures in units of the Boltzmann constant).
To this end, one may devise a dynamics having a temperature $T_{\rm target}$ as a dynamical
parameter that is designed to relax an initial
configuration of the system to canonical equilibrium at temperature $T_{\rm target}$, and then make the choice $T_{\rm
target}=T$. A~common practice is to employ a Langevin dynamics, i.e., a
\textit{noisy, dissipative} dynamics that mimics the interaction of the system with an
external heat bath at temperature $T_{\rm target}$ in terms of a
deterministic frictional force and an uncorrelated, Gaussian-distributed
random force added to the equation of motion \cite{Zwanzig:2001}. In
this approach, one then tunes suitably the strength of the random force such that the
Langevin dynamics relaxes at long times to canonical equilibrium at temperature
$T_{\rm target}$. The presence of dissipation renders the dynamics to be
\textit{irreversible in time}. A complementary approach to such a noisy, dissipative
dynamics was pioneered by Nos\'{e} and Hoover, in
which the dynamics is fully \textit{deterministic} and
\textit{time-reversible}, while achieving the same objective of relaxing
the system to canonical equilibrium at the desired temperature $T_{\rm
target}$ \cite{Nose:1984,Hoover:1985}; for a
review, see \cite{Morriss:1998,Klages:2007}.
 The time evolution under the
condition of relaxation at long times to canonical equilibrium at a
given temperature is said to represent isokinetic ensemble dynamics when taking place according to the Nos\'{e}--Hoover
equation of motion and to represent Langevin/canonical ensemble dynamics when taking place following the Langevin
equation of motion. 

 To illustrate in detail the distinguishing feature of the Nos\'{e}--Hoover vis-\`{a}-vis Langevin dynamics, consider an interacting $N$-particle system characterized by the set
$\{q_j,\pi_j\}$ of canonical coordinates and conjugated momenta. The
particles, which we take for simplicity to have the same mass $m$, interact with one another via the two-body interaction
potential $\Phi(\{q_j\})$. In the following, we consider $q_j$'s and
$\pi_j$'s to be one-dimensional variables for reasons of simplicity. Our analysis, however, extends
straightforwardly to higher dimensions. The Hamiltonian of the system is given by
\begin{equation}
{\cal H}_{\rm system}=\sum_{j=1}^N\frac{\pi_j^2}{2m}+\Phi(\{q_j\}),
\label{eq:H-bare}
\end{equation}
where the first term on the right-hand side stands for the kinetic
energy of the system. 

In the approach due to Langevin, the dynamical
equations of the system are given by
\begin{eqnarray}
&&\frac{{\rm d}q_j}{{\rm d}t}=\frac{\pi_j}{m},~~\frac{{\rm d}\pi_j}{{\rm
d}t}=-\gamma \frac{\pi_j}{m}-\frac{\partial
\Phi(\{q_j\})}{\partial q_j}+\eta_j(t), 
\label{eq:Langevin-eom}
\end{eqnarray}
where $t$ denotes time, $\gamma>0$ is the dissipation constant, while
$\eta_j(t)$ is a Gaussian, white noise satisfying 
\begin{equation}
\overline{\eta_j(t)}=0,~\overline{\eta_j(t)\eta_k(t')}=2D\delta_{jk}\delta(t-t').
\label{eq:Langevin-noise}
\end{equation}

Here, the overbars denote averaging over noise realizations,
while $D>0$ characterizes the strength of the noise. The dynamics
(\ref{eq:Langevin-eom}) are evidently not time-reversal invariant.
Choosing $D=\gamma T_{\rm target}$ ensures that the dynamics
(\ref{eq:Langevin-eom}) relaxes at long times to the canonical
distribution at $T_{\rm target}$ given by \cite{Zwanzig:2001}
\begin{equation}
P(\{q_j,\pi_j\})\propto \exp(-{\cal H}_{\rm system}/T_{\rm
target}),
\label{eq:canonical-distribution}
\end{equation} in which the kinetic energy density of the system
fluctuates around the average value $T_{\rm target}/2$.

In the approach due
to Nos\'{e} and Hoover, a degree of freedom $s$ augmenting the set $\{q_j,\pi_j\}$ is introduced, which is taken to
characterize an external heat reservoir that interacts with the
system through the momenta $\pi_j$'s. The Hamiltonian of the combined system is
given by
\begin{equation}
{\cal H}=\sum_{j=1}^N\frac{\pi_j^2}{2ms^2}+\Phi(\{q_j\})+\frac{p_s^2}{2Q}+(N+1)T_{\rm
target}
\ln s,
\label{eq:NH-H}
\end{equation}
where $Q$ is the mass and $p_s$ is the conjugated momentum of the
additional degree of freedom. The~dynamics of the system is given by the
following Hamilton equations of motion:
\begin{eqnarray}
&&\frac{{\rm d}q_j}{{\rm d}t}=\frac{\pi_j}{ms^2},~~\frac{{\rm d}\pi_j}{{\rm d}t}=-\frac{\partial
\Phi(\{q_j\})}{\partial q_j}, \nonumber \\
\label{eq:NH-eom} \\
&&\frac{{\rm d}s}{{\rm d}t}=\frac{p_s}{Q},~~\frac{{\rm d}p_s}{{\rm
d}t}=\sum_{j=1}^N
\frac{\pi_j^2}{ms^3}-(N+1)\frac{T_{\rm target}}{s}. \nonumber 
\end{eqnarray}
It may be easily checked that unlike dynamics (\ref{eq:Langevin-eom}),~dynamics (\ref{eq:NH-eom}) is invariant under time reversal. In terms of new variables

\begin{equation}
p_j\equiv\frac{\pi_j}{s},~~\zeta \equiv \frac{p_s}{Q},
\end{equation}
and rescaled time 
\begin{equation}
\tilde{t}\equiv \frac{t}{s},
\label{eq:time-rescaling}
\end{equation}
one obtains from the Hamilton Equations (\ref{eq:NH-eom}) the
~following dynamics:
\begin{eqnarray}
&&\frac{{\rm d}q_j}{{\rm d}\tilde{t}}=\frac{p_j}{m}, 
\label{eq:NH-dynamics1} \\
&&\frac{{\rm d}p_j}{{\rm d}\tilde{t}}=-\frac{\partial \Phi(\{q_j\})}{\partial
q_j}-\zeta p_j, \label{eq:NH-dynamics2} \\
&&\frac{{\rm d}s}{{\rm d}\tilde{t}}=\zeta s, \label{eq:NH-dynamics3} \\
&&\frac{{\rm d}\zeta}{{\rm d}\tilde{t}}=\frac{1}{Q}\left(\sum_{j=1}^N
\frac{p_j^2}{m}-(N+1)T_{\rm
target}\right)=\frac{1}{\tau^2}\Big(\frac{K(P)}{K_0}-1\Big),
\label{eq:NH-dynamics4}
\end{eqnarray}
where $K(P)\equiv \sum_{j=1}^N p_j^2/(2m)$ is the kinetic energy, while we have defined
\begin{equation}
K_0 \equiv (N+1)\frac{T_{\rm target}}{2},~~\tau^2\equiv
\frac{Q}{2K_0}.
\end{equation}

From Equations~(\ref{eq:NH-dynamics1})--(\ref{eq:NH-dynamics4}), we observe that a complete
description of the time evolution of the system is given in terms of
Equations~(\ref{eq:NH-dynamics1}), (\ref{eq:NH-dynamics2}), and
(\ref{eq:NH-dynamics4}), without any reference to Equation~(\ref{eq:NH-dynamics3}) for $s$, so~that, as far as
the description of the system is concerned, the variable $s$ is an
irrelevant one that may be ignored. We note in passing that a different, but closely related, Hamiltonian
giving directly the Nos\'{e}-Hoover equations of motion but without any
time scaling, as in Equation~(\ref{eq:time-rescaling}), is discussed in
\cite{Dettmann}.
We will from now on drop the tilde
over time in order not to overload the notation. Let us note that, in terms of the variables
$p_j$'s, the
Hamiltonian (\ref{eq:NH-H}) takes the form
\begin{equation}
{\cal H}=\sum_{j=1}^N\frac{p_j^2}{2m}+\Phi(\{q_j\})+\frac{Q\zeta^2}{2}+(N+1)T
\ln s.
\label{eq:NS-H-1}
\end{equation}

From Equation~(\ref{eq:NH-dynamics4}), we find that, in the stationary state
$({\rm d}\zeta/{\rm d}t=0$), the kinetic energy of the system equals $(N+1) T_{\rm
target}/2$ (the extra factor of unity
takes care of the presence of the additional degree of freedom $s$). For
large $N \gg 1$, we then have the desired result: an ensemble of initial
conditions under the evolution given by Equations~(\ref{eq:NH-dynamics1}), (\ref{eq:NH-dynamics2}), and
(\ref{eq:NH-dynamics4}) evolves at long times to a~stationary state in
which the average kinetic energy density has the value $T_{\rm target}/2$. The quantity $\tau$ in Equation~(\ref{eq:NH-dynamics4})
denotes a relaxation timescale over which the kinetic energy 
relaxes to its target value. Beyond the average kinetic energy, it has
been demonstrated by invoking the phase space continuity equation that the distribution 
\begin{equation}
f\propto \exp\left[-\left(\sum_{j=1}^N \frac{p_j^2}{2m}+
\Phi(\{q_j\})+Q\zeta^2/2\right)/T_{\rm
target}\right]
\label{eq:canonical-distribution-full-system}
\end{equation}
is a stationary
state of the Nos\'{e}--Hoover dynamics \cite{Hoover:1985}. It then follows that the corresponding stationary distribution for the
system variables $\{q_j,p_j\}$ is the canonical equilibrium
distribution:
\begin{equation}
P(\{q_j,p_j\})\propto \exp\left[-\left(\sum_{j=1}^N \frac{p_j^2}{2m}+
\Phi(\{q_j\})\right)/T_{\rm
target}\right],
\label{eq:canonical-distribution-1}
\end{equation}
normalized as $\int\left(\prod_{j=1}^N {\rm d}q_j {\rm
d}p_j\right)P(\{q_j,p_j\})=1$. Thus, the dynamics (\ref{eq:NH-dynamics1})--(\ref{eq:NH-dynamics4}) that includes the
additional dynamical variable $s$ nevertheless preseves the canonical
equilibrium distribution of the system.
A general formalism for constructing modified Hamiltonian dynamical
systems that preserve a canonical equilibrium distribution on adding a time evolution equation for
a single additional thermostat variable is discussed in \cite{Ramshaw}.

Equation~(\ref{eq:canonical-distribution-1}) implies that the single-particle momentum
distribution $P(p)$, defined such that $P(p){\rm d}p$
gives the probability that a randomly chosen particle has its momentum
between $p$ and $p+{\rm d}p$, is a Gaussian distribution with mean zero
and width equal to $T_{\rm target}$:
\begin{equation}
P(p)=\frac{1}{\sqrt{2\pi m T_{\rm target}}}\exp\left(-\frac{p^2}{2mT_{\rm
target}}\right).
\label{eq:canonical-distribution-2}
\end{equation}
Consequently, the moments $\langle p^n \rangle \equiv
\int_{-\infty}^\infty {\rm d}p~p^nP(p)$, with $n=1,2,3,\ldots$, satisfy $\langle p^4\rangle/\langle
p^2 \rangle^2=3$.

In the above backdrop, the principal objective of this work is to
answer the question: what is the effect of inter-particle
interactions on the relaxation properties of the Nos\'{e}--Hoover
dynamics?
More specifically, considering a system embedded in a $d$-dimensional
space, we ask: do systems with long-range interactions, in which the
inter-particle interaction decays slower than $1/r^d$, behave in a~similar way to short-range systems that have the inter-particle
interaction decaying faster than $1/r^d$? How does the timescale over
which the phase space distribution relaxes to its canonical equilibrium
form behave in the two cases, and, in particular, is there a system-size
dependence in the timescale for long-range systems with respect to short-range ones?
Studying these issues is particularly relevant and timely in the wake of
recent surge in interest across physics in long-range interacting (LRI)
systems. 

LRI systems may display a notably distinct thermodynamic behavior with
respect to short-range ones \cite{Campa:2009,Bouchet:2010,Campa:2014,Levin:2014,Gupta:2017-1}.
These systems are characterized by a two-body interaction potential
$V(r)$ that decays asymptotically with inter-particle separation $r$ as
$V(r)\sim r^{-\alpha}$, with $0\le \alpha \le d$ in $d$ spatial
dimensions. The limit $\alpha\to 0$ corresponds to the case of
mean--field interaction. Examples of LRI systems are self-gravitating systems, plasmas,
fluid dynamical systems, and some spin systems. One of the striking
dynamical features resulting from long-range interactions is the
occurrence of non-equilibrium quasi-stationary states (QSSs) during relaxation
of LRI systems towards equilibrium. These states have lifetimes that
diverge with the number of particles constituting the system, so that, in
the thermodynamic limit, the system remains trapped in QSSs and does not
attain equilibrium. Only for a~finite number of particles do the QSSs eventually evolve
towards equilibrium. Even in equilibrium, LRI systems may exhibit
features such as ensemble inequivalence and a negative heat capacity in the microcanonical
ensemble that are unusual for short-range systems.

In this work, we address our aforementioned queries within the ambit of a model system
comprising classical $XY$-spins occupying the sites of a one-dimensional
periodic lattice and interacting via a long-range (specifically, a
mean--field interaction in which every spin interacts with every other
and a short-range (specifically, a nearest-neighbor interaction in which
every spin interacts with its left and right neighbors) interaction.
With an aim to study the equilibrium properties as well as relaxation towards
equilibrium, we simulate the Nos\'{e}--Hoover dynamics of the model by
integrating the corresponding equations of motion in time. A signature
of canonical equilibrium is a single-particle momentum distribution that is
Gaussian (see Equation~(\ref{eq:canonical-distribution-2})). We find that the
equilibrium properties of our model system evolving under the
Nos\'{e}--Hoover dynamics coincide with those within the canonical ensemble. As regards relaxation towards canonical
equilibrium, we observe that starting from out-of-equilibrium initial
conditions, the average kinetic energy of the system
relaxes to its target canonical-equilibrium value over a {\it size-independent} timescale. However, quite
surprisingly, our results indicate that under the same conditions and with only
long-range interactions present in the system, the momentum distribution relaxes to its Gaussian form in equilibrium over a scale that {\it diverges} with the system size. On adding short-range interactions, the
relaxation is found to occur over a timescale that has a much weaker dependence on
system size. This system-size dependence vanishes when only short-range
interactions are present in the system. An implication of such an ultra-slow relaxation when
only long-range interactions are present in the system is that
macroscopic observables other than the average kinetic energy when
estimated in the Nos\'{e}--Hoover dynamics may take an unusually
long time to relax to its canonical equilibrium value. Our work underlines the
crucial role that interactions play in deciding the equivalence between
Nos\'{e}--Hoover and canonical equilibrium.

The paper is organized as follows. In Section \ref{sec:model}, we
describe the model of study. In Section \ref{sec:caloric-curve}, we
obtain the so-called caloric curve of the model within the canonical
ensemble, which we eventually invoke in later parts of the paper to
decide on the equivalence of the equilibrium properties of the
Nos\'{e}--Hoover dynamics and canonical equilibrium. In Section \ref{sec:results}, we present results from
simulations of the Nos\'{e}--Hoover dynamics of the model, and discuss the implications and
relevance of the results. The paper
ends with conclusions in Section \ref{sec:conclusions}.

\section{Model of Study}
\label{sec:model}
Our system of study comprises a one-dimensional periodic lattice of
$N$ sites. Each site of the lattice is occupied by a unit-inertia rotor characterized by its angular
coordinate $\theta_j \in [0,2\pi)$ and the corresponding conjugated
momentum $p_j$, with $j=1,2,\ldots,N$. One may also think of the rotors as
representing classical $XY$-spins. Note that both of the $\theta_j$'s and
the $p_j$'s are one-dimensional variables. There exist both a long-range (specifically, a global
or a mean--field) coupling and a short-range (specifically,
nearest-neighbor) coupling between the rotors. Thus, a rotor on site $j$
interacts with strength $J/(2N)$ with rotors on all the other sites and with
strength $K$ with the rotor occupying the $(j-1)$-th and
the $(j+1)$-th site. The
Hamiltonian of the system is given by \cite{Campa:2006,Dauxois:2010}
\begin{eqnarray}
{\cal H}=\sum_{j=1}^N\frac{p_j^2}{2}+\frac{J}{2N}\sum_{j,k=1}^N\left[1-\cos(\theta_j-\theta_k)\right]
+K\sum_{j=1}^N\left[1-\cos(\theta_{j+1}-\theta_j)\right];~~\theta_{N+1}\equiv
\theta_1,p_{N+1}\equiv p_1.
\label{eq:H} 
\end{eqnarray}

Note that, for $K=0$, the Hamiltonian (\ref{eq:H}) reduces to that of the
widely-studied Hamiltonian mean--field (HMF) model \cite{Antoni:1995}, which is regarded as
a paradigmatic model to study statics and dynamics of LRI systems \cite{Campa:2014}. On the other hand, for $J=0$, the model (\ref{eq:H})
reduces to a short-ranged $XY$ model in one dimension.

In the following, we take both the mean--field coupling $J$ and the
short-range coupling $K$ to be positive, thereby modeling ferromagnetic
global and nearest-neighbor couplings. Consequently, both the
long-range and the short-range coupling between the rotors favor an
ordered state in which all the
rotor angles are equal, thereby minimizing the potential energy
contribution to the total energy. Such~a~tendency is, however, opposed by
the kinetic energy contribution whose average in equilibrium may be characterized by a
temperature by invoking the Theorem of Equipartition. Noting that, for a~given $N$, the
total potential energy is bounded from above while the total kinetic
energy is not, one~expects the system to show in equilibrium an
ordered/magnetized phase at low
energies/temperatures and a disordered/unmagnetized phase at high
energies/temperatures. This~scenario holds
even with $K=0$. 

The amount of order in
the system is characterized by the $XY$ magnetization
\begin{equation}
{\bf m}\equiv \frac{1}{N}\left(\sum_{j=1}^N \cos \theta_j,\sum_{j=1}^N
\sin \theta_j\right),
\end{equation}
which is a vector whose length $m$ has the thermodynamic value in
equilibrium denoted by $m^{\rm eq}$ that is nonzero in the ordered phase and zero in the disordered phase.
For $K=0$, the corresponding HMF model is known to display a
second-order phase transition between a high-temperature unmagnetized
phase and a low-temperature magnetized phase at the critical
temperature $T_c =J/2$, with the corresponding critical energy density
being $u_c=3J/4$ \cite{Campa:2014}. On the other hand, invoking the
Landau's argument for the absence of any phase transition at a finite
temperature in a one-dimensional model with only short-range
interactions, one may conclude for $J=0$ that the corresponding short-ranged $XY$ model does not display any phase
transitions, though it has been shown to have interesting dynamical
effects \cite{Escande:1994}. For general $J \ne 0,K\ne0$, when both
long-range and short-range interactions are present, the model displays a
second-order phase transition between an ordered and
a disordered phase \cite{Campa:2006,Dauxois:2010}. Note that all the mentioned phase transitions are
continuous. Although ensemble equivalence is not guaranteed for LRI
systems, it has been argued that inequivalence arises when one has
a first-order phase transition in the canonical ensemble, and not when one has a second-order transition \cite{Bouchet:2005-1}. Consequently, we may regard the phase diagram of model (\ref{eq:H}) to
be equivalent within microcanonical and canonical ensembles. For an explicit demonstration of ensemble equivalence for the model
(\ref{eq:H}), one may refer to 
\cite{Dauxois:2010}. 

In the following
section, we will obtain the caloric curve of model (\ref{eq:H}) that
relates the equilibrium internal energy with the equilibrium temperature
of the system.

\section{The Caloric Curve within the Canonical Ensemble}
\label{sec:caloric-curve}
As mentioned in the preceding section, model (\ref{eq:H}) is known
to have equivalent microcanonical and canonical ensemble descriptions in
equilibrium. Consequently, in obtaining the caloric curve of the model,
which will be invoked to decide the equivalence between the equilibrium
properties of the Nos\'{e}--Hoover dynamics and canonical equilibrium, it will suffice to restrict our analysis to the canonical ensemble
description of the model.

The Langevin/canonical ensemble dynamics (\ref{eq:Langevin-eom}) for the model (\ref{eq:H})
comprises the set of time-evolution
equations
\begin{eqnarray}
&&\frac{{\rm d}\theta_j}{{\rm d}t}=p_j, \nonumber \\
\label{eq:H-eom-canonical-ensemble} \\
&&\frac{{\rm d}p_j}{{\rm d}t}=-\gamma p_j+\frac{J}{N}\sum_{k=1}^N
\sin(\theta_k-\theta_j)+K\left[\sin(\theta_{j+1}-\theta_j)+\sin(\theta_{j-1}-\theta_j)\right]+\eta_j(t), \nonumber 
\end{eqnarray}
with the properties of the noise $\eta_j(t)$ given by Equation~(\ref{eq:Langevin-noise}) with $D=\gamma T$.
Within the microcanonical ensemble description of the
system, the time evolution of the variables $\{\theta_j,p_j\}$ is given
by Hamilton equations obtained from
Equation~(\ref{eq:H-eom-canonical-ensemble}) by setting $\gamma$ to zero. The
Nos\'{e}--Hoover dynamics of the variables $\{\theta_j,p_j\}$ is obtained from Equations~(\ref{eq:NH-dynamics1}) and (\ref{eq:NH-dynamics2}) as
\begin{eqnarray}
&&\frac{{\rm d}\theta_j}{{\rm d}t}=p_j,\nonumber \\
\label{eq:eom-our-model} \\
&&\frac{{\rm d}p_j}{{\rm
d}t}=\frac{J}{N}\sum_{k=1}^{N}\sin(\theta_k-\theta_j)+K\left[\sin(\theta_{j+1}-\theta_j)+\sin(\theta_{j-1}-\theta_j)\right]-\zeta
p_j,\nonumber 
\end{eqnarray}
where the time evolution of the variable $\zeta$ is given by Equation~(\ref{eq:NH-dynamics4}). 

In order to derive the desired caloric curve of model (\ref{eq:H})
within the canonical ensemble,
we start with the canonical partition function of the system at temperature $T$ given
by \linebreak $Z_N\equiv\int\left(\prod_{j=1}^N {\rm d}\theta_j{\rm d}p_j\right) \exp[-\beta
{\cal H}(\{\theta_j,p_j\})]$, with $\beta\equiv 1/T$. Using Equation~(\ref{eq:H}),
we get
\begin{equation}
Z_N=\left(\frac{2\pi}{\beta}\right)^{N/2}e^{-\beta JN/2-\beta KN}\int\left(\prod_{j=1}^N {\rm d}\theta_j\right)
\exp\Big[
\frac{\beta
J}{2N}\left\{\Big(\sum_{j=1}^N\cos\theta_j\Big)^2+\Big(\sum_{j=1}^N\sin\theta_j\Big)^2\right\}
+\beta K\sum_{j=1}^N\cos(\theta_{j+1}-\theta_j)\Big].
\label{eq:canonical-partition-function}
\end{equation}

Using the Hubbard--Stratonovich transformation $\exp(ax^2)=1/(\sqrt{4\pi
a})\int_{-\infty}^\infty {\rm
d}z~\exp\left(-\frac{z^2}{4a}+ zx\right), {a>0}$ in Equation~(\ref{eq:canonical-partition-function}), we obtain
\begin{eqnarray}
&&Z_N=\left(\frac{2\pi}{\beta}\right)^{N/2}e^{-\beta JN/2-\beta
KN}\frac{N\beta J}{2\pi}\int_{-\infty}^\infty {\rm d}z_1\int_{-\infty}^\infty
{\rm d}z_2 \int \left(\prod_{j=1}^N {\rm d}\theta_j\right)
\exp\Big[-\frac{N\beta J}{2}(z_1^2+z_2^2)\nonumber \\
&&+\beta J z_1
\sum_{j=1}^N\cos \theta_j +\beta J z_2\sum_{j=1}^N\sin
\theta_j+\beta
K\sum_{j=1}^N\cos(\theta_{j+1}-\theta_j)\Big].
\label{eq:canonical-partition-function-auxiliary-fields}
\end{eqnarray}

Writing $z_1=z\cos \phi, z_2=z\sin \phi$, with real
$z=(z_1^2+z_2^2)^{1/2}>0$ and
$\phi \in [0,2\pi)$ given by $\phi=\tan^{-1}(z_2/z_1)$, we get
\begin{eqnarray}
&&Z_N=\left(\frac{2\pi}{\beta}\right)^{N/2}e^{-\beta JN/2-\beta
KN}\frac{N\beta J}{2\pi}\int_0^{2\pi} {\rm d}\phi\int_0^\infty 
{\rm d}z~z \int \left(\prod_{j=1}^N {\rm d}\theta_j\right)
\exp\Big[-\frac{N\beta J}{2}z^2\nonumber \\
&&+\beta J z\sum_{j=1}^N\cos (\theta_j-\phi)+\beta
K\sum_{j=1}^N\cos(\theta_{j+1}-\theta_j)\Big].
\label{eq:canonical-partition-function-auxiliary-fields-again}
\end{eqnarray}

In view of the invariance of the Hamiltonian (\ref{eq:H}) under
rotation by an equal amount of all the $\theta_j$'s, we get \cite{Gupta:2017}
\begin{equation}
Z_N=\left(\frac{2\pi}{\beta}\right)^{N/2}e^{-\beta JN/2-\beta
KN}N\beta J
\int_0^\infty{\rm d}z~ z\int \left(\prod_{j=1}^N {\rm d}\theta_j\right)
\exp\Big[-\frac{N\beta J}{2}z^2+\beta J z \sum_{j=1}^N\cos \theta_j+\beta
K\sum_{j=1}^N\cos(\theta_{j+1}-\theta_j)\Big].
\label{eq:canonical-partition-function-auxiliary-fields-nophi}
\end{equation}

In order to proceed further, we consider separately the
cases $K=0$ and $K\ne 0$ in the following.

\subsection{$K=0$}
\label{sec:Keq0}
For $K=0$, Equation~(\ref{eq:canonical-partition-function-auxiliary-fields-nophi}) yields
\begin{equation}
Z_N=\left(\frac{2\pi}{\beta}\right)^{N/2} N \beta J\int_0^\infty 
{\rm d}z~z\exp\left[-N\left\{\frac{\beta J}{2}(1+z^2)-\ln \left(\int_0^{2\pi} {\rm d}\theta
~\exp(\beta J z \cos \theta)\right)\right\}\right]. 
\label{eq:ZN-K0-explicit}
\end{equation}

In the thermodynamic limit, $Z_N$ may be approximated by
invoking the saddle-point method to perform
the integration in $z$ on the right-hand side; one gets
\begin{equation}
Z_N=\left(\frac{2\pi}{\beta}\right)^{N/2}N\beta J z_s\exp\left[-N\left\{\frac{\beta J}{2}(1+z_s^2)-\ln \left(\int_0^{2\pi} {\rm d}\theta
~\exp(\beta J z_s \cos \theta)\right)\right\}\right], 
\label{eq:ZN-K0-explicit-1}
\end{equation}
where the saddle-point value $z_s$ solves the equation
\begin{equation}
z_s=\frac{I_1(\beta J z_s)}{I_0(\beta J z_s)},
\label{eq:saddle-point-K0}
\end{equation}
with $I_n(x)=(1/(2\pi))\int_0^{2\pi}{\rm d}\theta~\exp(x\cos
\theta)\cos(n\theta)$ being the modified Bessel function of first kind and
of order $n$. It may be shown by following the arguments given in \cite{Gupta:2017} that $z_s$ is nothing but the stationary
magnetization $m^{\rm eq}$. Equation~(\ref{eq:saddle-point-K0}) has a
trivial solution $m^{\rm eq}=0$ valid at all temperatures, while a non-zero
solution exists for $\beta \ge \beta_c=2/J$ \cite{Campa:2014}. In fact,
the system shows a continuous transition, from a magnetized
phase ($m^{\rm eq}\ne 0$) at low temperatures to an unmagnetized
phase ($m^{\rm eq}=0$) at high temperatures at the critical temperature
$T_c=J/2$ \cite{Campa:2014}. 

In the thermodynamic limit, the internal energy density of the system \linebreak
$u=-\lim_{N \to \infty}(1/N){\rm d} \ln Z_N/{\rm d} \beta$
is obtained by using Equations~(\ref{eq:ZN-K0-explicit-1}) and (\ref{eq:saddle-point-K0}) as
\begin{equation}
u=\frac{1}{2\beta}+\frac{J}{2}\left(1-(m^{\rm eq})^2\right);~m^{\rm
eq}=\frac{I_1(\beta J m^{\rm eq})}{I_0(\beta J m^{\rm eq})},
\label{eq:internal-energy-long}
\end{equation}
yielding the critical energy density
\begin{equation}
u_c=\frac{3J}{4}.
\end{equation}

Equation~(\ref{eq:internal-energy-long}) gives the caloric curve of the
model (\ref{eq:H}) at canonical equilibrium for $J \ne 0, K=0$.

\subsection{$K\ne 0$}
\label{sec:Kne0}
For $K \ne 0$, Equation~(\ref{eq:canonical-partition-function-auxiliary-fields-nophi}) gives
\begin{eqnarray}
&&Z_N=\left(\frac{2\pi}{\beta}\right)^{N/2}N\beta J\int_0^\infty 
{\rm d}z~z\exp\left[-\frac{N\beta J}{2}(1+z^2)-\beta KN\right]{\cal
Z}_N,\label{eq:canonical-partition-function-3}\\
&&{\cal Z}_N\equiv\int \left(\prod_{j=1}^N {\rm d}\theta_j\right)
\exp\left[\beta J z \sum_{j=1}^N\cos
\theta_j+\beta
K\sum_{j=1}^N\cos(\theta_{j+1}-\theta_j)\right],
\label{eq:calZN-definition}
\end{eqnarray}
where we may identify the factor ${\cal Z}_N$ with the canonical partition function
of a $1d$ periodic chain of $N$ interacting angle-only rotors, where a rotor on each site interacts with strength $K$ with
the rotor on the
left nearest-neighbor and the right nearest-neighbor site, and also with
an external field of strength $J z$ along the $x$
direction.

One may evaluate ${\cal Z}_N$ by rewriting it in terms of a
transfer operator $T(\theta,\theta')$ as
\begin{eqnarray}
&&{\cal Z}_N=\int \left(\prod_{j=1}^N {\rm d}\theta_j\right){\cal
T}(\theta_1,\theta_2){\cal T}(\theta_2,\theta_3)\ldots{\cal
T}(\theta_N,\theta_1), \\
&&{\cal T}(\theta_j,\theta_{j+1})\equiv \exp\left[\beta J z
\left\{\frac{\cos \theta_j+\cos \theta_{j+1}}{2}\right\}+\beta
K\cos(\theta_{j+1}-\theta_j)\right].
\end{eqnarray}

Let $\{\lambda_m\}$ denote the set of eigenvalues of the transfer
operator ${\cal T}(\theta,\theta')$. In other words, denoting the
eigenfunctions of ${\cal T}(\theta,\theta')$ as $f_m(\theta)$, we have $\int {\rm d}\theta'~{\cal T}(\theta,\theta')f_m(\theta')=\lambda_m
f_m(\theta)$.
In terms of $\{\lambda_m\}$, we~obtain 
\begin{equation}
{\cal Z}_N=\sum_m \left[\lambda_m\left(\beta Jz,\beta K\right)\right]^N.
\label{eq:canonical-partition-function-lambda}
\end{equation}

For large $N$, the sum in Equation~(\ref{eq:canonical-partition-function-lambda})
is dominated by the largest eigenvalue $\lambda_{\rm
max}=\lambda_{\rm max}\left(\beta Jz,\beta K\right)$,
yielding  
\begin{equation}
{\cal
Z}_N=\lambda_{\rm max}^N.
\label{eq:canonical-partition-function-lambdamax}
\end{equation}

Substituting Equation~(\ref{eq:canonical-partition-function-lambdamax})
in Equation~(\ref{eq:canonical-partition-function-3}), and approximating the
integral on the right-hand side of the latter by the saddle-point method, one gets
\begin{equation}
Z_N=\left(\frac{2\pi}{\beta}\right)^{N/2}N\beta
Jz_s\exp\left[-N\left\{\frac{\beta J}{2}(1+z_s^2)+\beta K-\ln \lambda_{\rm
max}\left(\beta J
z_s,\beta K\right)\right\}\right],
\label{eq:canonical-partition-function-saddle-point}
\end{equation}
where $z_s$ solves the saddle-point equation $z_s\equiv\sup_z
\widetilde{\phi}(\beta,z)$, with $\widetilde{\phi}(\beta,z)$ being the free-energy function:
\begin{equation}
-\widetilde{\phi}(\beta,z)\equiv-\frac{1}{2}\ln \beta-\frac{\beta
J}{2}(1+z^2)-\beta K+\ln
\lambda_{\rm max}\left(\beta J
z,\beta K\right). 
\label{eq:saddle-point-definition-2}
\end{equation}

The saddle-point equation may thus be written as 
\begin{equation}
z_s=\frac{\partial \ln \lambda_{\rm max}\left(\beta J z,\beta
K\right)}{\partial(\beta Jz)}\Big|_{z=z_s}.
\label{eq:saddle-point-equation}
\end{equation}

Equation~(\ref{eq:canonical-partition-function-saddle-point}) gives the dimensionless free energy per
rotor, $\phi(\beta)\equiv -\lim_{N \to \infty} (\ln Z_N)/N$, as~$-\phi(\beta)=\sup_z \left[-\widetilde{\phi}(\beta,z)\right]$, where we have suppressed the dependence of $\phi(\beta)$ on $K$. We~thus~have 
\begin{equation}
-\phi(\beta)\equiv-\frac{1}{2}\ln \beta-\frac{\beta J}{2}(1+z_s^2)-\beta K+\ln \lambda_{\rm
max}\left(\beta J
z_s,\beta K\right).
\label{eq:phi-beta-final}
\end{equation}
\newpage
Note that the free energy at a given temperature has a definite value
given by Equation~(\ref{eq:phi-beta-final}), and~is obtained by substituting the saddle-point solution $z_s$ into the
expression for the free-energy function~$\widetilde{\phi}(\beta,z)$. 

In the thermodynamic limit, the internal energy density of the system
\linebreak$u=-\lim_{N \to \infty}(1/N){\rm d} \ln Z_N/{\rm d} \beta$
is obtained as
\begin{equation}
u=\frac{1}{2\beta}+\frac{J}{2}(1+z_s^2)+\beta J z_s\frac{{\rm
d}
z_s}{{\rm d}\beta}+K-\frac{{\rm d} \ln \lambda_{\rm max}(\beta J
z,\beta K)}{{\rm d} \beta}\Big|_{z=z_s}.
\end{equation}

Using Equation~(\ref{eq:saddle-point-equation}), and the fact that, as for $K=0$, the quantity $z_s$ is nothing but the stationary
magnetization $m^{\rm eq}$, we get
\begin{equation}
u=\frac{1}{2\beta}+\frac{J}{2}\left(1-(m^{\rm eq})^2\right)+\beta
Jm^{\rm eq}\frac{{\rm d}m^{\rm eq}}{{\rm d}\beta}+K-K\frac{\partial \ln
\lambda_{\rm max}(\beta J m^{\rm eq},\beta K)}{\partial (\beta K)},
\label{eq:internal-energy}
\end{equation}
with $m^{\rm eq}$ satisfying
\begin{equation}
m^{\rm eq}=\frac{\partial \ln \lambda_{\rm max}\left(\beta Jz,\beta
K\right)}{\partial(\beta J z)}\Big|_{z=m^{\rm eq}}.
\label{eq:saddle-point-equation-mst}
\end{equation}

To proceed, we need to find $\lambda_{\rm max}(\beta Jz,\beta K)$. We
consider separately the cases $J=0$ and $J \ne 0$. 

\subsubsection{$J=0$}
\label{sec:Kne0Jeq0}
In this case, it may be easily checked
that the eigenvalues of ${\cal T}$ are given by $2\pi I_m(\beta K)$
with the corresponding eigenvector given by plane waves
$\exp(iq\theta)/\sqrt{2\pi}$ \cite{Dauxois:2010}. Using $I_0(x)>I_1(x)>I_2(x)\ldots$, we conclude that $\lambda_{\rm
max}(0,\beta K)=I_0(\beta K)$. Equation~(\ref{eq:saddle-point-equation-mst}) then yields $m^{\rm eq}=0$, while
Equation~(\ref{eq:internal-energy}) gives
\begin{equation}
u=\frac{1}{2\beta}+K\left(1-\frac{I_1(\beta K)}{I_0(\beta K)}\right),
\label{eq:internal-energy-J0}
\end{equation}
where we have used the result ${\rm d}I_0(x)/{\rm d}x=I_1(x)$.
Equation~(\ref{eq:internal-energy-J0}) is the desired caloric curve of
the model (\ref{eq:H}) within the
canonical ensemble for $J=0, K\ne 0$.

\subsubsection{$J \ne 0$}
\label{sec:Kne0Jne0}
In this case, not knowing the analytic forms of the eigenvalues of ${\cal T}$, we resort to a numerical scheme to estimate the largest eigenvalue
$\lambda_{\rm max}(\beta Jz,\beta K)$. To this end, we discretize the angles over the interval
$[0,2\pi)$ as $\theta_j^{(a_j)}=a_j\Delta \theta$, with
$a_j=1,2,\ldots,P$ and $\Delta \theta=2\pi/P$ for any large positive
integer $P$ (we choose $P=30$). The operator ${\cal T}(\theta,\theta')$ then takes the form of a
matrix of size $P \times P$, whose largest eigenvalue may be estimated numerically by employing the so-called power method
\cite{Larson:2017}. Noting that ${\cal T}(\theta,\theta')$ is a finite-dimensional real square matrix with positive entries, the
application of the Perron--Frobenius theorem implies the existence of its largest eigenvalue that is real and non-degenerate. 
At given values of $T,K,J,z$, once $\lambda_{\rm max}(\beta Jz,\beta K)$
has been estimated numerically, we compute the free-energy function
$\widetilde{\phi}(\beta,z)$ as a function of $z$ by using
Equation~(\ref{eq:saddle-point-definition-2}). We then find numerically the value of $z$ at which
the computed free-energy function attains its minimum value. As
discussed above, this minimizer is the equilibrium
magnetization of the system at the given values of $T,K,J$. In order to
obtain the caloric curve, one has to estimate numerically the derivative
$\partial \ln \lambda_{\rm max}(\beta J m^{\rm eq},\beta K)/\partial
(\beta K)$, and then use Equation~(\ref{eq:internal-energy}).

\section{Results and Discussion}
\label{sec:results}
In this section, we discuss the results on equilibrium as well as
relaxation properties of model~(\ref{eq:H}) obtained by performing numerical integration of the
Nos\'{e}--Hoover equations of motion~(\ref{eq:eom-our-model}).
The numerical integration involved using a
fourth-order Runge--Kutta method with timestep ${\rm d}t=0.01$. 

\subsection{Results in Equilibrium}

\begin{figure}[!ht]
\begin{center}
\includegraphics[width=160mm]{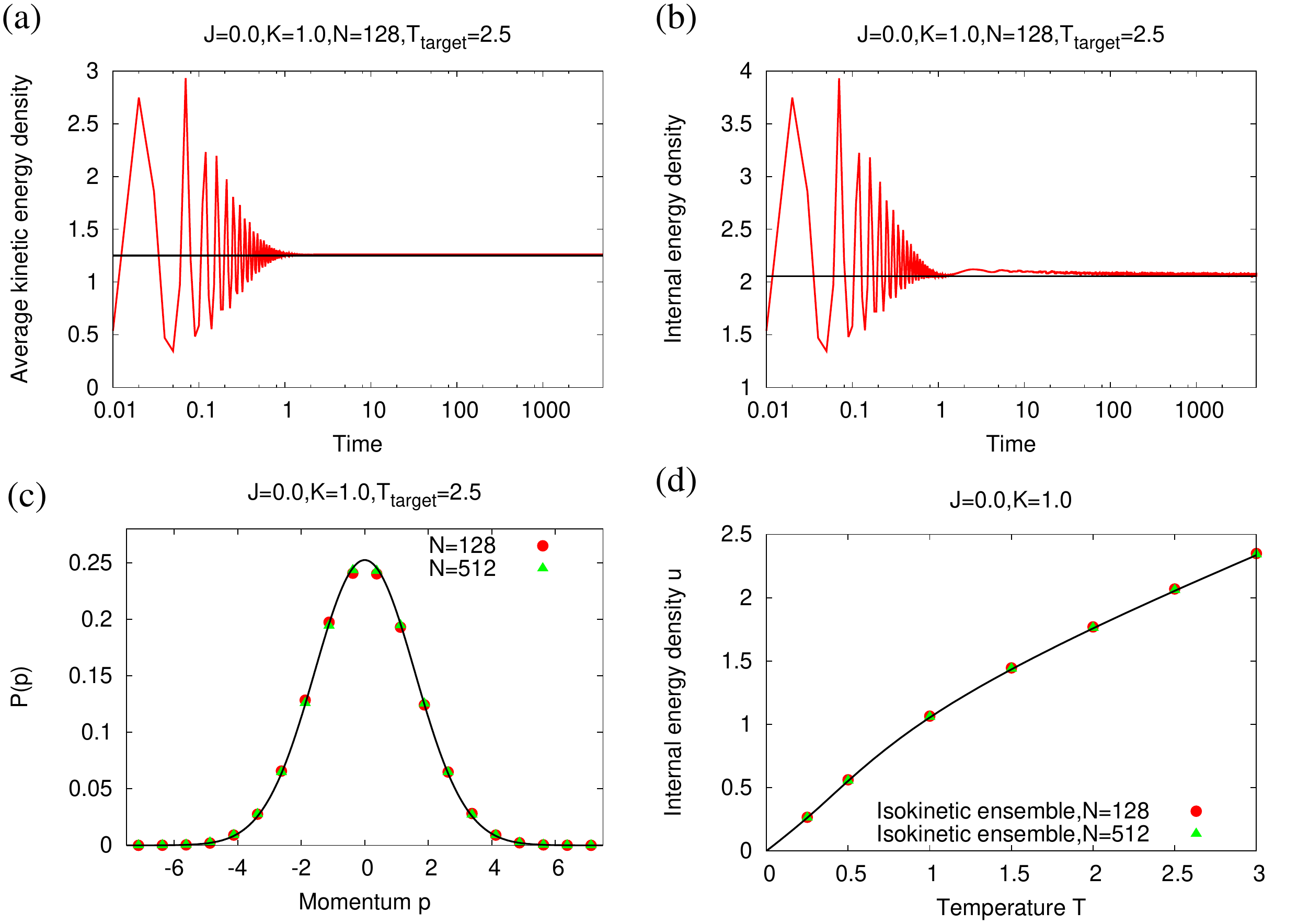}
\end{center}\vspace{-6pt}
\caption{Comparison of Nos\'{e}--Hoover and canonical equilibrium results for
model (\ref{eq:H}) with $J=0.0,K=1.0$ (that is, with only short-range
interactions). (\textbf{a}) variation of the average kinetic energy
density with time. The black line denotes the value $T_{\rm
target}/2$; (\textbf{b}) variation of the internal energy
density with time. The black line denotes the average internal
energy density within the canonical ensemble given by
Equation~(\ref{eq:internal-energy-J0}); (\textbf{c}) stationary
single-particle momentum
distribution obtained from momentum values measured at time $t=5000$. The black line
denotes a Gaussian distribution with zero mean and width equal to
$T_{\rm target}$; (\textbf{d}) caloric curve for two system sizes,
$N=128$ and $N=512$. The~black line shows the caloric
curve within the canonical ensemble given by
Equation~(\ref{eq:internal-energy-J0}). The~data for the Nos\'{e}--Hoover dynamics are generated by
integrating the equations of motion (\ref{eq:eom-our-model}) using a~fourth-order
Runge--Kutta method with timestep equal to $0.01$. The initial condition
corresponds to the $\theta_j$'s independently and uniformly distributed
in $[0,2\pi)$ and the $p_j$'s independently sampled from a~Gaussian
distribution with zero mean and width equal to $0.5$. The initial value
of the parameter $\zeta$ is $2$, while we have taken $\tau=0.01$.}
\label{fig:short-only}
\end{figure}

\begin{figure}[!ht]
\begin{center}
\includegraphics[width=160mm]{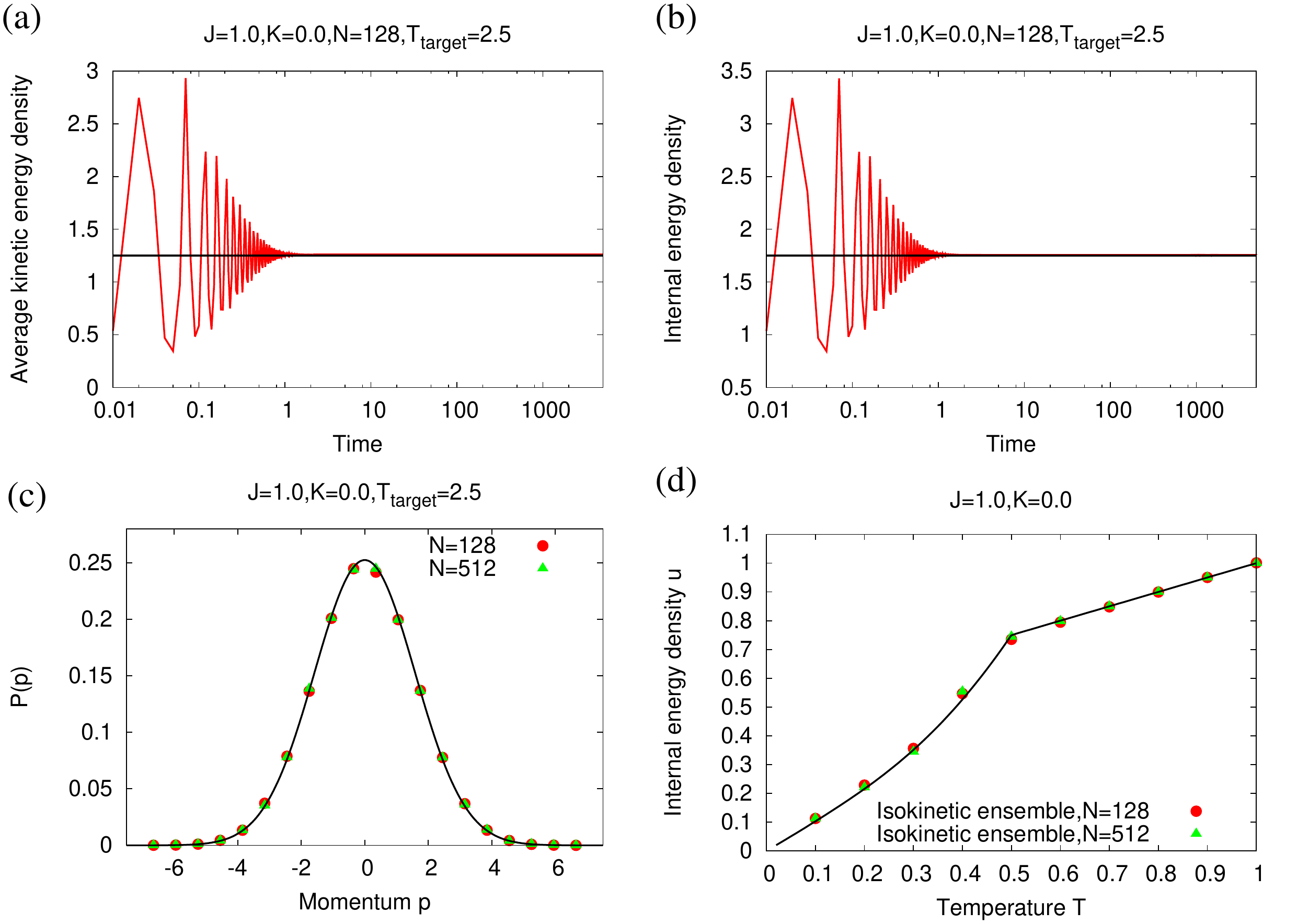}
\end{center}\vspace{-6pt}
\caption{Comparison of Nos\'{e}--Hoover and canonical equilibrium results for the
model (\ref{eq:H}) with $J=1.0,K=0.0$ (that is, with only long-range
interactions); (\textbf{a}) variation of the average kinetic energy
density with time. The black line denotes the value $T_{\rm
target}/2$; (\textbf{b}) variation of the internal energy
density with time. The black line denotes the average internal
energy density within the canonical ensemble given by
Equation~(\ref{eq:internal-energy-long}); (\textbf{c}) stationary
single-particle momentum
distribution obtained from momentum values measured at time $t=5000$. The black line
denotes a Gaussian distribution with zero mean and width equal to
$T_{\rm target}$; (\textbf{d}) caloric curve for two system sizes,
$N=128$ and $N=512$. The black line shows the caloric
curve within the canonical ensemble given by
Equation~(\ref{eq:internal-energy-long}). The~data for the Nos\'{e}--Hoover
dynamics are generated by
integrating the equations of motion (\ref{eq:eom-our-model}) using a~fourth-order
Runge--Kutta method with timestep equal to $0.01$. The initial condition
corresponds to the $\theta_j$'s independently and uniformly distributed
in $[0,2\pi)$ and the $p_j$'s independently sampled from a~Gaussian
distribution with zero mean and width equal to $0.5$. The initial value
of the parameter $\zeta$ is $2$, while we have taken $\tau=0.01$.}
\label{fig:long-only}
\end{figure}

\begin{figure}[!ht]
\begin{center}
\includegraphics[width=150mm]{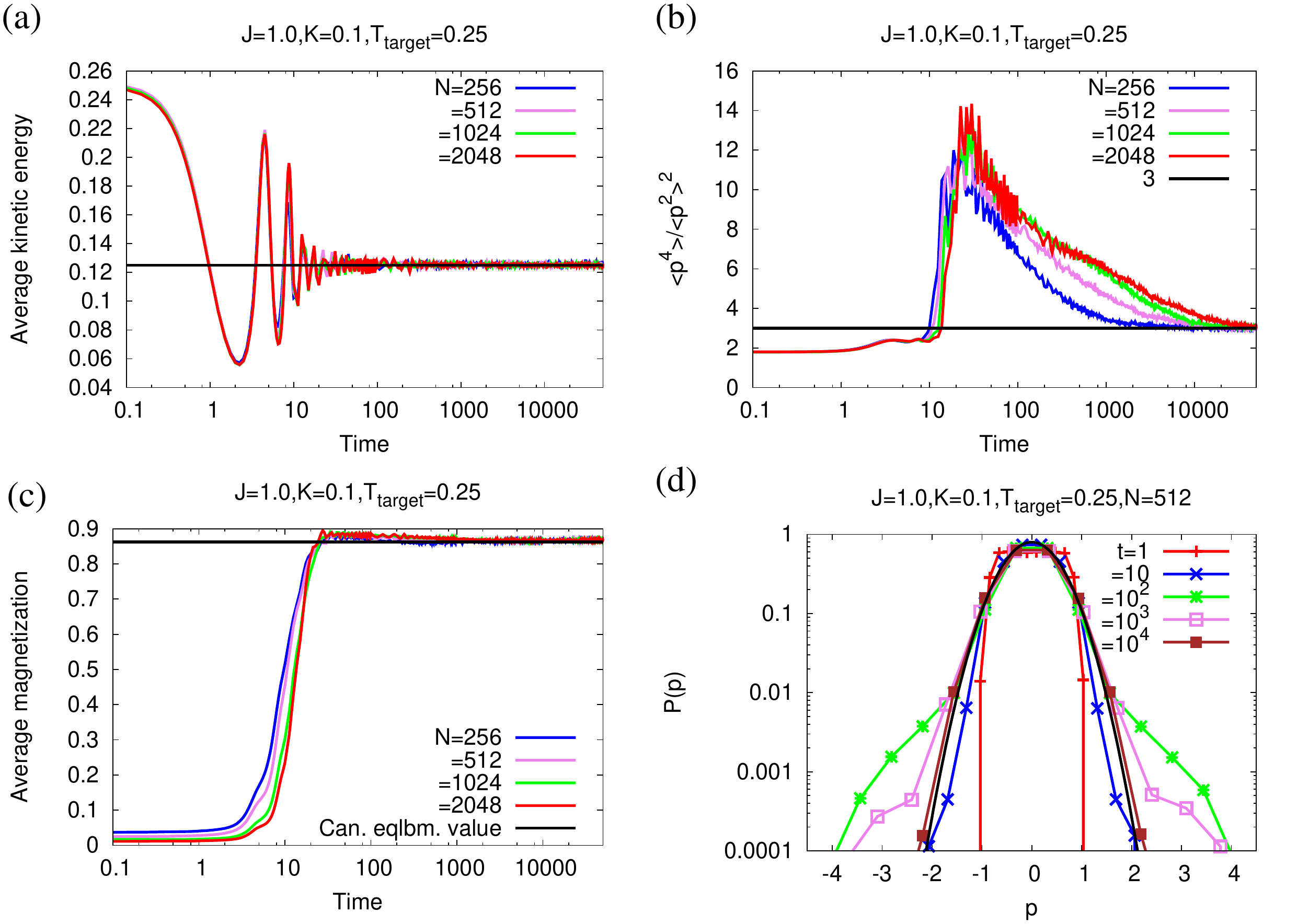}
\end{center}
\caption{Relaxation properties of the Nos\'{e}--Hoover dynamics for model (\ref{eq:H}) with $J=1.0,K=0.1$. (\textbf{a}) variation of the average kinetic energy
density with time, for four different system sizes. The black line denotes the value $T_{\rm
target}/2$; (\textbf{b}) variation of the ratio
$\langle p^4 \rangle/\langle p^2 \rangle^2$ with time, for four
different system sizes. The black line denotes the value $3$
corresponding to a Gaussian distribution; (\textbf{c}) variation of the
magnetization with time, again for four different system sizes. The
black line denotes the canonical equilibriu m value obtained by the
method described in Section \ref{sec:Kne0Jne0}; (\textbf{d})
single-particle momentum
distribution as a function of time, for system size $N=512$. The black
line denotes a Gaussian distribution with zero mean and width equal to
$T_{\rm target}$, Equation~(\ref{eq:canonical-distribution-2}). The data for the Nos\'{e}--Hoover dynamics are generated by
integrating the equations of motion (\ref{eq:eom-our-model}) using a fourth-order
Runge--Kutta method with timestep equal to $0.01$. The initial condition
corresponds to the $\theta_j$'s independently and uniformly distributed
in $[0,2\pi)$ and the $p_j$'s independently and uniformly distributed in
$[-\sqrt{1.5},\sqrt{1.5}]$. The initial value
of the parameter $\zeta$ is $2$, while we have taken $\tau=1.0$.}
\label{fig:J1-K0-1}
\end{figure}

Here, we discuss the Nos\'{e}--Hoover equilibrium properties for
model (\ref{eq:H}). The initial condition corresponds to the $\theta_j$'s independently and uniformly distributed
in $[0,2\pi)$ and the $p_j$'s independently sampled from a Gaussian
distribution with zero mean and width equal to $0.5$. The initial value
of the parameter $\zeta$ is $2.0$, while we have taken $\tau=0.01$. In
Figure~\ref{fig:short-only}, we consider the case when only long-range
interactions are present in system ($J=1.0,K=0.0$). Figure~\ref{fig:short-only}a 
shows
for $T_{\rm target}=2.5$
that the average kinetic energy relaxes at long times to the value
$T_{\rm target}/2$, as desired. Figure~\ref{fig:short-only}b shows for the same value of $T_{\rm target}$ that the average
internal energy has the same value in the stationary state as the one in
canonical equilibrium given by Equation~(\ref{eq:internal-energy-long}); Figure~\ref{fig:short-only}c shows the single-particle momentum
distribution $P(p)$ in the stationary state. We observe that $P(p)$ has the correct canonical-equilibrium form of a Gaussian
distribution, which further corroborates the property of the
Nos\'{e}--Hoover dynamics that the canonical distribution
(\ref{eq:canonical-distribution-1}) is a stationary state of the dynamics.
Figure~\ref{fig:short-only}d shows for a range of values of the temperature $T=T_{\rm target}$ that the caloric curve obtained within the Nos\'{e}--Hoover
dynamics in equilibrium coincides with that within the canonical
ensemble given by Equation~(\ref{eq:internal-energy-long}). Figure~\ref{fig:short-only}a--c
refer to the system size $N=128$, while Figure~\ref{fig:short-only}d refers to two system
sizes, namely, $N=128$ and $N=512$. The aforementioned observed
properties of the Nos\'{e}--Hoover dynamics have been checked to hold for
(i) the case when only short-range interactions are
present in the system (see Figure~\ref{fig:long-only} that corresponds to $J=0.0, K=1.0$), in which case the caloric curve within the
canonical ensemble is given by Equation~(\ref{eq:internal-energy-J0}), and
(ii) when both long- and short-range interactions are present in the
system (data not shown; see, however, Figure~\ref{fig:J1-K0-1}c).

\subsection{Results out of Equilibrium}

Here, we discuss the relaxation properties of the Nos\'{e}--Hoover
dynamics for model (\ref{eq:H}). The~initial condition corresponds
to the so-called water-bag distribution that has both $\theta$ and $p$ uniformly distributed over given intervals \cite{Campa:2014}.
We consider $\theta_j$'s to be independently and uniformly distributed
in $[0,2\pi)$ and the $p_j$'s to be independently and uniformly distributed in
$[-\sqrt{1.5},\sqrt{1.5}]$. The initial value
of the parameter $\zeta$ is $2.0$, while we have taken $\tau=1.0$.

\begin{figure}[!ht]
\begin{center}
\includegraphics[width=150mm]{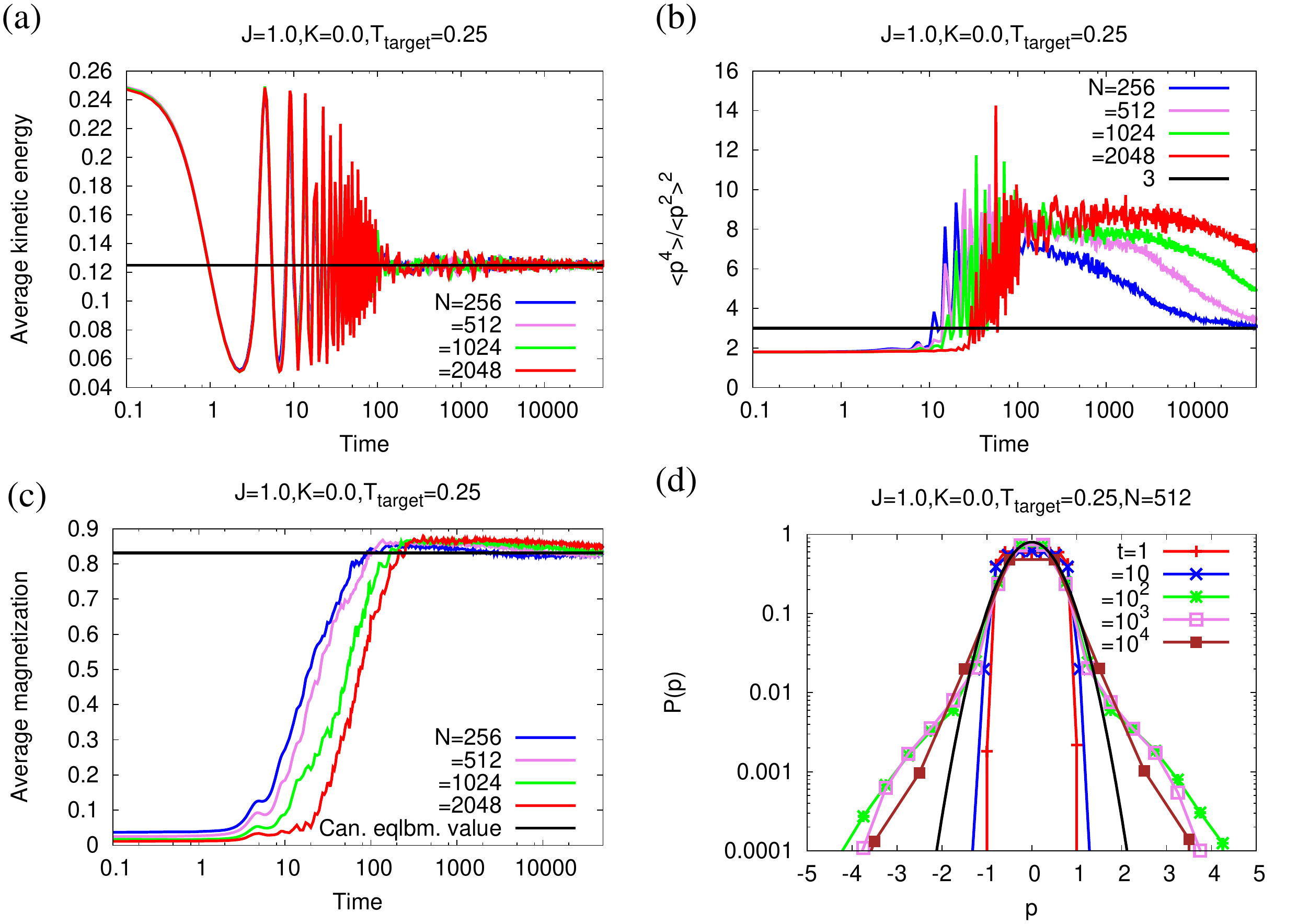}
\end{center}
\caption{Relaxation properties of the Nos\'{e}--Hoover dynamics for the
model (\ref{eq:H}) with $J=1.0,K=0.0$ (that is, with only long-range
interactions). (\textbf{a}) variation of the average kinetic energy
density with time, for four different system sizes. The black line denotes the value $T_{\rm
target}/2$; (\textbf{b}) variation of the ratio
$\langle p^4 \rangle/\langle p^2 \rangle^2$ with time, for four
different system sizes. The black line denotes the value $3$
corresponding to a Gaussian distribution; (\textbf{c}) variation of the
magnetization with time, again for four different system sizes. The
black line denotes the canonical equilibrium value given by
Equation~(\ref{eq:saddle-point-K0}); (\textbf{d})
single-particle momentum
distribution as a function of time, for system size $N=512$. The black
line denotes a Gaussian distribution with zero mean and width equal to
$T_{\rm target}$, Equation~(\ref{eq:canonical-distribution-2}). The~data for the Nos\'{e}--Hoover dynamics are generated by
integrating the equations of motion (\ref{eq:eom-our-model}) using a~fourth-order
Runge--Kutta method with timestep equal to $0.01$. The initial condition
corresponds to the $\theta_j$'s independently and uniformly distributed
in $[0,2\pi)$ and the $p_j$'s independently and uniformly distributed in
$[-\sqrt{1.5},\sqrt{1.5}]$. The initial value
of the parameter $\zeta$ is $2$, while we have taken $\tau=1.0$.}
\label{fig:J1-K0}\vspace{-8pt}
\end{figure}

Let us start with a discussion of the results in Figure~\ref{fig:J1-K0}
that corresponds to the case when only long-range interactions are
present in the system (\ref{eq:H}).
In Figure~\ref{fig:J1-K0}a, we see that, for four different system sizes, the
average kinetic energy density relaxes at long times to the target 
value $T_{\rm target}/2$ over a timescale that \textit{does not depend
on the system size}. A Gaussian distribution for the momentum, expected
in canonical equilibrium, is characterized by a value $3$ of the ratio
$\langle p^4\rangle/\langle p^2 \rangle^2$ (see Equation~(\ref{eq:canonical-distribution-2})). We see in Figure~\ref{fig:J1-K0}b
that, in contrast to Figure~\ref{fig:J1-K0}a, this ratio, however, relaxes to the
canonical equilibrium value over a time that \textit{depends on the
system size}, and which grows with increase of $N$. Figure~\ref{fig:J1-K0}c shows
that the long-time magnetization value reached by the Nos\'{e}--Hoover
dynamics coincides with the canonical equilibrium value for all system
sizes. On the basis of these results, we conclude that, with
only long-range interactions in system (\ref{eq:H}), only the second
moment of the momentum distribution relaxes to its canonical equilibrium
value over a size-independent timescale, while higher moments (and
consequently, the whole of the momentum distribution) relax to their
canonical equilibrium values over a time that grows with the system
size. The latter fact is demonstrated in Figure~\ref{fig:J1-K0}d that shows
for $N=512$ the time evolution of the single-particle momentum distribution.
\vspace{6pt}

\begin{figure}[!ht]
\begin{center}
\includegraphics[width=150mm]{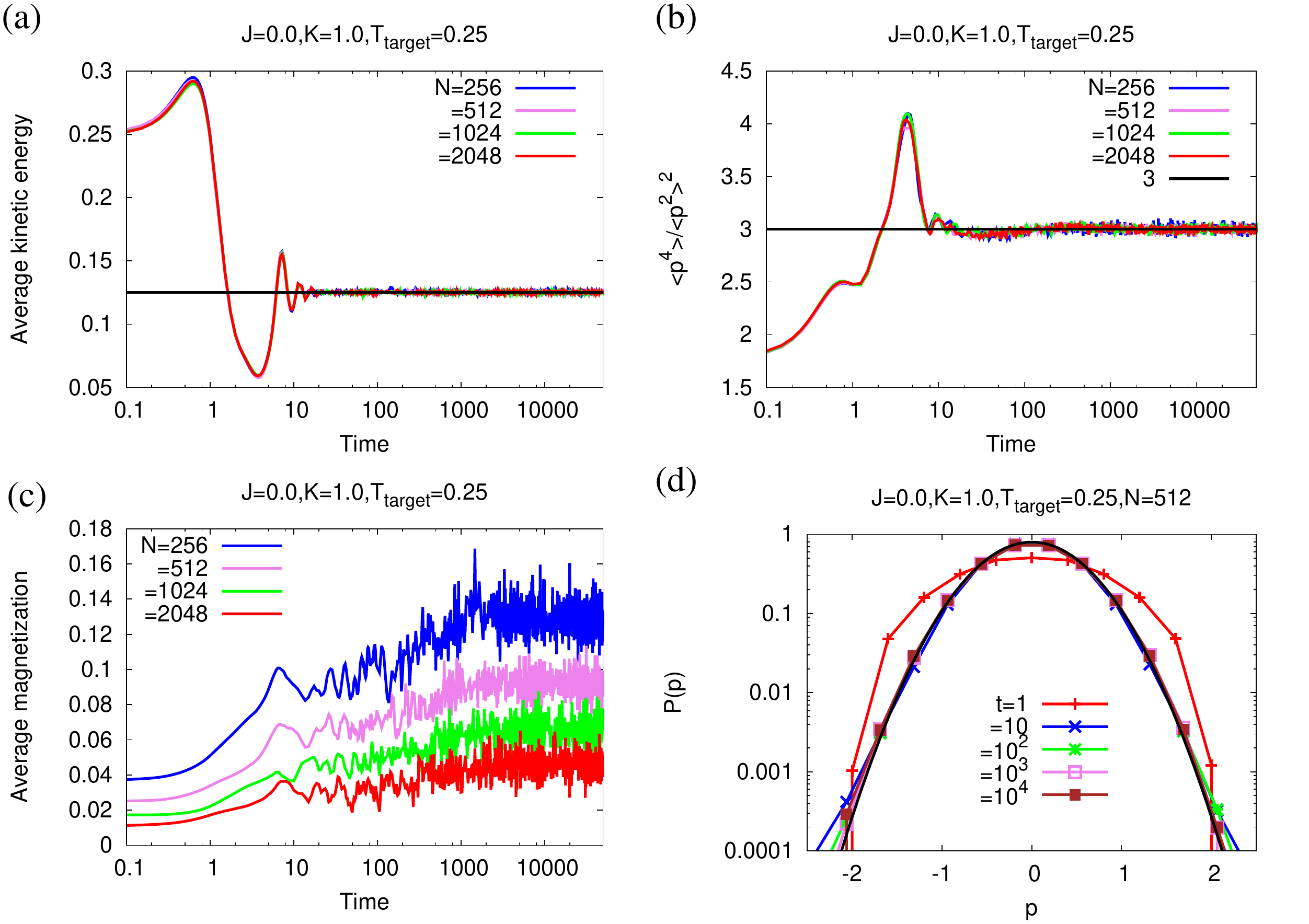}
\end{center}
\caption{Relaxation properties of the Nos\'{e}--Hoover dynamics for the
model (\ref{eq:H}) with $J=0.0,K=1.0$ (that is, with only short-range
interactions). (\textbf{a}) variation of the average kinetic energy
density with time, for four different system sizes. The black line denotes the value $T_{\rm
target}/2$; (\textbf{b}) variation of the ratio
$\langle p^4 \rangle/\langle p^2 \rangle^2$ with time, for four
different system sizes. The black line denotes the value $3$
corresponding to a Gaussian distribution; (\textbf{c}) variation of the
magnetization with time, again for four different system sizes. The equilibrium magnetization goes to zero with
increase of $N$ as $m^{\rm eq} \sim 1/\sqrt{N}$; (\textbf{d})
single-particle momentum
distribution as a function of time, for system size $N=512$. The black
line denotes a Gaussian distribution with zero mean and width equal to
$T_{\rm target}$, Equation~(\ref{eq:canonical-distribution-2}). The~data for the Nos\'{e}--Hoover dynamics are generated by
integrating the equations of motion (\ref{eq:eom-our-model}) using a~fourth-order
Runge--Kutta method with timestep equal to $0.01$. The initial condition
corresponds to the $\theta_j$'s independently and uniformly distributed
in $[0,2\pi)$ and the $p_j$'s independently and uniformly distributed in
$[-\sqrt{1.5},\sqrt{1.5}]$. The initial value
of the parameter $\zeta$ is $2$, while we have taken $\tau=1.0$.}
\label{fig:J0-K1}
\end{figure}

The feature of a size-independent timescale for the relaxation of the
average kinetic energy density to its canonical equilibrium value,
observed in the case of purely long-range interactions in model~(\ref{eq:H}), also holds on adding short-range interactions to the model
and when the latter are the only interactions present in the system (see
Figures~\ref{fig:J1-K0-1}a and \ref{fig:J0-K1}a).
Moreover, in all cases, the long-time value of the magnetization matches
with its canonical equilibrium value (see
Figures~\ref{fig:J1-K0-1}c and \ref{fig:J0-K1}c). The
most significant difference in the relaxation properties that is
observed on adding short-range interactions may be inferred by comparing
Figure~\ref{fig:J1-K0-1}b and Figure~\ref{fig:J1-K0}b: the
very strong size-dependence observed in the relaxation of the 
ratio $\langle p^4 \rangle/\langle p^2 \rangle^2$ to its canonical
equilibrium value gets substantially weakened on adding short-range
interactions with coupling strength as low as $K=0.1$ compared to the
value of the long-range coupling constant $J=1.0$. Similar inference may
be drawn from a comparison of Figure~\ref{fig:J1-K0-1}d and Figure~\ref{fig:J1-K0}d. This observation has
an immediate and an important implication: additional short-range
interactions speed up the relaxation of the momentum distribution
towards canonical equilibrium. The aforementioned system-size dependence
vanishes on turning off long-range interactions, so that the only
remnant interactions in the system are the short-range ones (see 
Figure~\ref{fig:J0-K1}b,d).

\vspace{6pt}

\section{Conclusions}
\label{sec:conclusions}
In this paper, we investigated the relaxation properties of the Nos\'{e}--Hoover
dynamics of many-body interacting Hamiltonian systems, with an emphasis on the effect of inter-particle
interactions. The dynamics aim to generate the canonical equilibrium
distribution of a system at the desired temperature by employing time-reversible, deterministic dynamics. To pursue our study, we considered a representative model
comprising $N$ classical $XY$-spins occupying the sites of a one-dimensional
periodic lattice. The spins interact with one another via both a
long-range interaction, modelled as a mean--field interaction in which
every spin interacts with every other, and~a~short-range one, modelled
as a nearest-neighbor interaction in which every spin interacts with its
left and right neighboring spins. We studied the Nos\'{e}--Hoover
dynamics of the model through $N$-body integration of the corresponding
equations of motion. Canonical
equilibrium is characterized by a momentum distribution that is
Gaussian. We found that the
equilibrium properties of our model system evolving according to
Nos\'{e}--Hoover dynamics are in excellent agreement with exact analytic results for the
equilibrium properties derived within the canonical ensemble. Moreover,
while starting from out-of-equilibrium initial conditions, the average kinetic energy of the system
relaxes to its target value over a {\it size-independent} timescale. However, quite unexpectedly, we found that under the same conditions and with only
long-range interactions present in the system, the momentum distribution relaxes to its Gaussian form in equilibrium over a scale that {\it grows} with $N$. The $N$-dependence gets weaker on adding short-range
interactions, and vanishes when the latter are the only inter-particle
interactions present in the system. 

Viewed from the perspective of LRI
systems, the slow relaxation observed within the Nos\'{e}--Hoover
dynamics allows for drawing an analogy with a similar slow relaxation observed within the
microcanonical dynamics of isolated LRI systems, a phenomenon that leads
to the occurrence of nonequilibrium quasistationary states (QSSs) that have
lifetimes diverging with the system size~\cite{Yamaguchi:2004,Campa:2014}. Within a kinetic theory
approach, the QSSs are understood as stable, stationary solutions of the
so-called Vlasov equation that governs the time evolution of the
single-particle phase space distribution. The Vlasov equation is obtained as the
first equation of the Bogoliubov--Born--Green--Yvon--Kirkwood (BBGKY)
hierarchy by neglecting the correlation between particle trajectories, with corrections that
decrease with an increase of $N$. For large but finite $N$, the eventual
relaxation of QSSs towards equilibrium is understood as arising due
to these finite-$N$ corrections, the so-called collisional terms, to the
Vlasov equation. In models in which the momentum variables are one-dimensional, it has been shown by analyzing the behavior of
the dominant collisional term that Vlasov-stable phase-space distributions that are homogeneous in the coordinates evolve
on times much larger than $N$, thereby leading for the distributions to
characterize QSSs that have lifetimes diverging with $N$
\cite{Campa:2009,Levin:2014,Bouchet:2005-2}. In light of the foregoing discussions, it is
evidently pertinent and of immediate interest
to invoke a kinetic theory approach and investigate in the context of
the Nos\'{e}--Hoover dynamics of long-range
systems whether additional short-range interactions play the role
of collisional dynamics that speed up the relaxation of the system
towards canonical equilibrium. Work in this direction is in progress and
will be reported~elsewhere.

The agreement reported in this paper in the value of the average kinetic energy computed in
canonical equilibrium and within the Nos\'{e}--Hoover dynamics is
reminiscent of a similar agreement in the large-system limit between ensemble and time averages
predicted by Khinchin for the so-called sum-functions, that is,
functions such as the kinetic energy that are sums of
single-particle contributions~\cite{Khinchin:1949}. The result was
obtained for rarefied gases, which was later observed to also hold
for systems with
short-range interactions \cite{Mazur:1963,Livi:1987}. Our work hints at the validity of such a result
even for long-range systems, as is evident from the agreement in the
value of the average kinetic energy computed within the Nos\'{e}--Hoover
dynamics and in canonical equilibrium (see Figure~\ref{fig:J1-K0}a). This point warrants a more detailed
investigation that will be left for future studies.

\vspace{6pt} 

\section{Acknowledgments}The authors
thank W. G. Hoover for fruitful email exchanges. Thanks are due to Leticia
Cugliandolo for pointing out that a relaxation phenomenon similar to what is
reported in this paper, namely, a fast relaxation of the kinetic energy and a
much slower one of other observables, has also been observed in
spin-glass systems.



\begin{thebibliography}{999}
\bibitem{Zwanzig:2001}Zwanzig, R. {\em Nonequilibrium Statistical
Mechanics}; Oxford University Press: Oxford, UK, 2001.

\bibitem{Nose:1984}Nos\'{e}, S. A unified formulation of the constant
temperature molecular-dynamics methods. {\em J. Chem. Phys.} {\bf
1984}, {\em 81}, 511--519.

\bibitem{Hoover:1985}Hoover, W.G. Canonical dynamics: Equilibrium
phase-space distributions. {\em Phys. Rev. A} {\bf 1985}, {\em 31}, 1695--1697.

\bibitem{Morriss:1998}Morriss, G.P.;  Dettmann, C.P. Thermostats:
Analysis and application. {\em Chaos} {\bf 1998}, {\em 8}, 321--336.

\bibitem{Klages:2007}Klages, R. {\em Microscopic Chaos, Fractals and
Transport in Nonequilibrium Statistical Mechanics, monograph, Advanced
Series in Nonlinear Dynamics Vol. 24}; World Scientific, Singapore,
2007.

\bibitem{Dettmann}Dettmann C. P. and Morriss G. P. Hamiltonian
reformulation and pairing of Lyapunov exponents for Nos\'{e}-Hoover
dynamics. {\em Phys. Rev. E} {\bf 1997}, {\bf 55}, 3693--3696.

\bibitem{Ramshaw}Ramshaw J. D. General formalism for singly thermostated
Hamiltonian dynamics. {\em Phys. Rev. E} {\bf 2015}, {\em 92},
052138--052143.


\bibitem{Campa:2009}Campa, A.; Dauxois, T.; Ruffo, S. Statistical mechanics and dynamics of solvable models with
long-range interactions. {\em Phys. Rep.}
{\bf 2009}, {\em 480}, 57--159.

\bibitem{Bouchet:2010}Bouchet, F.; Gupta, S.; Mukamel, D. Thermodynamics
and dynamics of systems with long-range interactions. {\em Physica
A} {\bf 2010}, {\em 389}, 4389--4405. 

\bibitem{Campa:2014}Campa, A.; Dauxois, T.; Fanelli, D.; Ruffo, S. {\em
Physics of Long-Range Interacting Systems}; Oxford University Press: Oxford, UK, 2014.

\bibitem{Levin:2014}Levin, Y.; Pakter, R.; Rizzato, F.B.; Teles, T.N.; 
Benetti, F.P.C. Nonequilibrium statistical mechanics of systems with
long-range interactions. {\em Phys. Rep.} {\bf 2014}, {\em 535}, 1--60.

\bibitem{Gupta:2017-1}Gupta, S.;  Ruffo, S. The world of long-range
interactions: A bird’s eye view. {\em Int. J. Mod. Phys. A}
{\bf 2017}, {\em 32}, 1741018.

\bibitem{Campa:2006}Campa, A.; Giansanti, A.; Mukamel, D.; Ruffo, S.
Dynamics and thermodynamics of rotators interacting with both long- and
short-range couplings. {\em Physica A} {\bf 2006}, {\em 365}, 120--127.

\bibitem{Dauxois:2010}Dauxois, T.; de Buyl, P.; Lori, L.; Ruffo, S. Models with short- and long-range interactions: The phase diagram and
the reentrant phase. {\em J. Stat. Mech. Theory Exp.} {\bf 2010}, {\em 2010}, P06015.

\bibitem{Antoni:1995}Antoni, M.; Ruffo, S. Clustering and relaxation in
Hamiltonian long-range dynamics. {\em Phys. Rev. E} {\bf 1995}, {\em 52},~2361--2373.

\bibitem{Escande:1994}Escande, D.; Kantz, H.; Livi, R.; Ruffo, S. Self-consistent check of the
validity of Gibbs calculus using dynamical variables. {\em J. Stat.
Phys.} {\bf 1994}, {\em 76}, 605--626.

\bibitem{Bouchet:2005-1}Bouchet, F.; Barr\'{e}, J. Classification of
phase transitions and ensemble inequivalence, in systems with long range
interactions. {\em J. Stat. Phys.} {\bf 2005}, {\em 118}, 1073--1105.

\bibitem{Gupta:2017}Gupta, S. Spontaneous collective synchronization in
the Kuramoto model with additional non-local interactions. {\em J. Phys.
A: Math. Theor.} {\bf 2017}, {\em 50}, 424001.

\bibitem{Larson:2017}Larson, R. {\em Elementary Linear Algebra},  8th
ed.; Cengage Learning: Boston, MA, USA, 2017. A FORTRAN90 library that implements the power
method and is distributed under the GNU Lesser General Public License
(GPL) is available at \url{http://people.sc.fsu.edu/~jburkardt/f_src/power_method/power_method.html)}.

\bibitem{Yamaguchi:2004}Yamaguchi, Y.Y.; Barr\'{e}, J.; Bouchet, F.;
Dauxois, T.; Ruffo, S. Stability criteria of the Vlasov equation and
quasi-stationary states of the HMF model. {\em Physica A} {\bf 2004}, {\em 337}, 36--66.

\bibitem{Bouchet:2005-2}Bouchet, F.; Dauxois, T. Prediction of anomalous diffusion and algebraic relaxations for long-range interacting systems,
using classical statistical mechanics. {\em Phys. Rev. E} {\bf 2005},
{\em 72}, 045103.

\bibitem{Khinchin:1949}Khinchin, A.I. {\em Mathematical Foundations of
Statistical Mechanics}; Dover: New York, NY, USA, 1949.

\bibitem{Mazur:1963}Mazur, P.; van der Linden, J. Asymptotic form of
the structure function for real systems. {\em J. Math. Phys.} {\bf
1963}, {\em 4}, 271--277.

\bibitem{Livi:1987}Livi, R.; Pettini, M.; Ruffo, S.; Vulpiani, A. Chaotic
behavior in nonlinear Hamiltonian systems and equilibrium statistical
mechanics. {\em J. Stat. Phys.} {\bf 1987}, {\em 48}, 539--559.

\end{thebibliography}
\end{document}